\documentclass{article}  
\usepackage{spadre2008}
\usepackage{graphicx}
\frompage{000} \topage{000}                                              
\def\rightmark{Understanding Quarkonia at PHENIX}
\def\leftmark{M.J. Leitch (for the PEHNIX Collaboration)}
 
\title{Progress Towards Understanding Quarkonia at PHENIX} 
\authors{
{M.J. Leitch$^1$ {\it et al.} (for the PHENIX Collaboration)%
}\\[2.812mm]
{\normalsize
\hspace*{-8pt}$^1$ P-25, Los Alamos National Laboratory, MS H846, \\ 
Los Alamos, NM 87545\\[0.2ex] 
}}
 
\abstract{Quarkonia (J/$\psi$, $\psi^\prime$, $\chi_C$, $\Upsilon$) production
provides a sensitive probe of gluon distributions and their modification
in nuclei; and is a leading probe of the hot-dense (deconfined) matter
created in high-energy collisions of heavy ions. We will discuss the physics
of quarkonia production in the context of recent $p+p$ measurements at
PHENIX. We next discuss Cold-Nuclear Matter (CNM) effects as seen in our
measurements in $d+Au$ collisions - both for intrinsic physics such
as gluon saturation and final-state dissociation, and as a baseline
for studies in nucleus-nucleus collisions. Then we review the latest
nucleus-nucleus results in the light of the expected CNM effects, and
discuss two leading scenarios for the observed suppression patterns.
Finally we show the latest data from PHENIX, including new $d+Au$ data
from the 2007-2008 run; and then look into the future.}

\keyword{quarkonia, heavy-quarks, quark-gluon plasma, gluon saturation, cold nuclear matter.} 
\PACS{24.85.+p, 13.85.-t, 25.75.-q, 25.75.Nq}
 
\begin{document}
 
\maketitle
\setcounter{page}{1}

\section{Introduction}\label{intro}

\begin{figure}[htb]
\center
\includegraphics[width=0.55\textwidth,clip=]{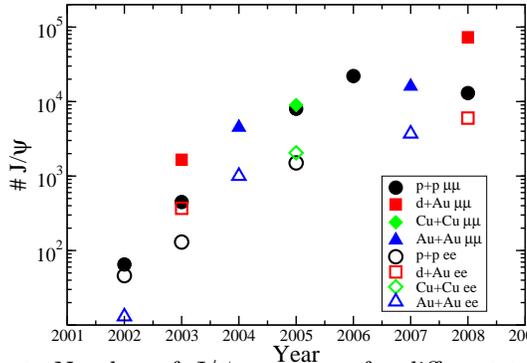}
\vspace*{-0.6cm}
\caption[]{Approximate Number of $J/\psi$s per year for different types
of collisions at PHENIX. Close symbols are for dimuons at forward rapidity,
and open symbols are for dielectrons at mid rapidity.}
\label{jsi_history_log}
\vspace*{-0.4cm}
\end{figure}

We discuss our present understanding of Quarkonia
($J/\psi$, $\psi^\prime$, $\chi_C$, $\Upsilon$) based on the measurements by PHENIX at RHIC.
We discuss 1) production, 2) cold nuclear matter (CNM) effects, 3) the effect
of the Quark Gluon Plasma (QGP), and then comment on prospects for the
future as RHIC luminosities increase and detector upgrades are installed.
As shown in Figure \ref{jsi_history_log}, the numbers of $J/\psi$ obtained in recent runs
has increased dramatically, with over 70,000 in the just completed $d+Au$ run.

\section{How are Quarkonia Produced}\label{production}  

Quarkonia are produced primarily via gluon-fusion, but it has proven
difficult for theoretical predictions to reproduce both
the cross section and the polarization of the $J/\psi$. The configuration
of the initially produced $c\bar{c}$ state remains unclear, and casts
uncertainty on what CNM effects it will experience in nuclei.
NRQCD models produce a $c\bar{c}$ in a color-octet state and are able to
reproduce the cross section, but predict large transverse polarization
at large $p_T$ - unlike the data from E866/NuSea\cite{e866_ups_pol} and
CDF\cite{cdf_polarization} which show
only small longitudinal polarization. However, a recent color-singlet model\cite{lansberg}
claims good agreement for both cross section and polarization.

Another complication in quarkonia production, particularly for the $J/\psi$,
is that about $\sim{40\%}$ of the $J/\psi$s come from decays of higher
mass resonances, namely the $\psi^\prime$ and $\chi_C$. Until recently, these fractions
have been inferred from measurements at other
energies. Now PHENIX has started to quantify these itself with initial
results indicating $8.6 \pm 2.5\%$ from the $\psi^\prime$ and $< 42\%$ from
the $\chi_C$. Another PHENIX measurement\cite{moreno} shows that $4 {{+3}\atop{-2}}\%$
of the $J/\psi$s come from decays of B-mesons, a contribution which is
strongest at larger $p_T$.

\section{What Cold Nuclear Matter (CNM) Effects are Important}\label{CNM}  

For Quarkonia produced in nuclei, e.g. in $p+A$ or $d+A$ collisions, several interesting
effects - usually called cold nuclear matter (CNM) effects, can occur. These include
modifications of the initial gluon density either according to traditional nuclear shadowing
models\cite{EKS,NDSG} that involve fits to deep-inelastic scattering and other data, or gluon saturation
models\cite{CGC}. In addition the initial-state projectile gluon may lose energy before it interacts
to form a $J/\psi$. Both of these effects can cause suppression of the produced $J/\psi$s
per nucleon-nucleon collision at large rapidity (or small x) relative to that observed
in p+p collisions. Finally, the $J/\psi$s can be suppressed by dissociation
of the $c{\bar{c}}$ by the nuclear medium in the final state.

\begin{figure}[htb]
\begin{minipage}[t]{0.55\linewidth}
  \includegraphics[width=\textwidth,clip=]{figures/alpha_x2xf.eps}
  \vspace*{-1.0cm}
  \caption[]{Nuclear dependence of $J/\psi$ production for three different
  energies vs $x_2$ and $x_F$. Where $x_F = x_1 - x_2$ and $x_1$ and $x_2$ are
  the momentum fractions in $d$ and $Au$ respectively.
  $\alpha$ is a representation of the nuclear dependence in terms
  of a power law, i.e. $\sigma_A = \sigma_N A^{\alpha}$.}
  \label{alpha_x2xf}
\end{minipage}
\hspace{0cm}
\begin{minipage}[t]{0.45\linewidth}
  \includegraphics[width=\textwidth,clip=]{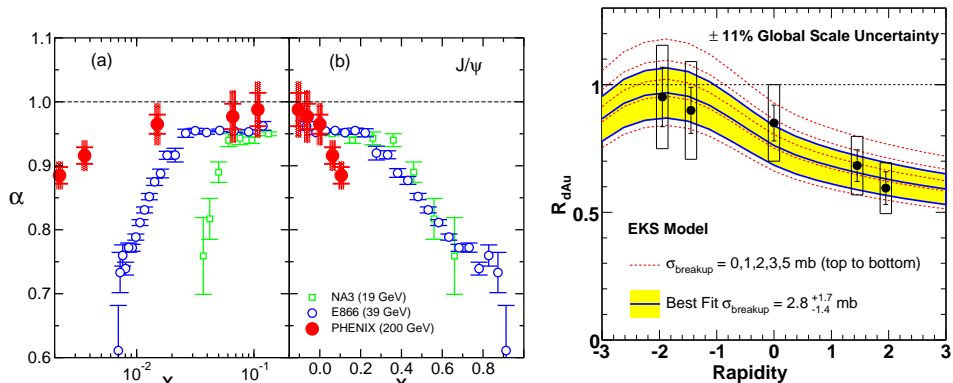}
  \vspace*{-1.1cm}
  \caption[]{Nuclear modification factor versus rapidity for $d+Au$ collisions.
  The yellow band bordered by black lines represents a fit to a model that contains
  EKS\cite{EKS} shadowing and a dissociation cross section.}
  \label{rdau_eksmodel}
\end{minipage}
\end{figure} 

A new analysis of the 2003 PHENIX $d+Au$ data, along with the new 2005 baseline $p+p$ data
have been put together to produce new nuclear modification factors for CNM\cite{ppg078},
as shown in Figure \ref{alpha_x2xf}, where they are compared to similar data at
lower energies. The lack of scaling with $x_2$ shown in the left panel of the figure
suggests that traditional shadowing models, which should have a universal $x_2$ dependence,
are not the dominant physics. The approximate scaling with $x_F$ (right panel), at least
for the lower energy data that extends to large $x_F$, hints that initial-state energy loss
or gluon saturation may be the dominant physics.

\begin{figure}[htb]
\begin{minipage}[t]{0.50\linewidth}
  \includegraphics[width=\textwidth,clip=]{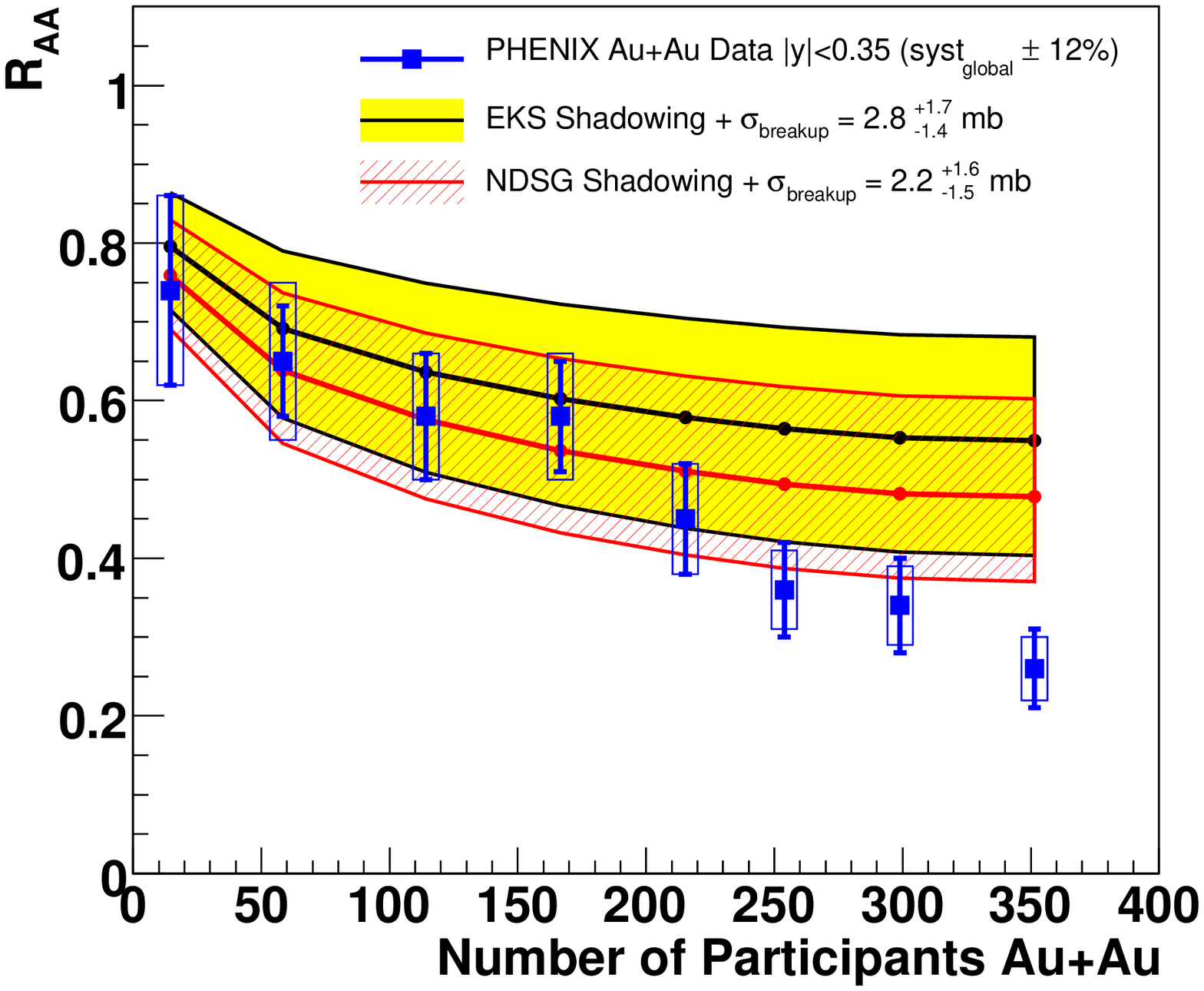}
  \vspace*{-1.2cm}
  \caption[]{Extrapolation of the simple CNM model shown in
  Figure \ref{rdau_eksmodel} to Au+Au collisions at mid rapidity. Results for
  both EKS\cite{EKS} and NDSG\cite{NDSG} shadowing are shown.}
  \label{figure_rauau_project_mid}
\end{minipage}
\hspace{0cm}
\begin{minipage}[t]{0.50\linewidth}
  \includegraphics[width=\textwidth,clip=]{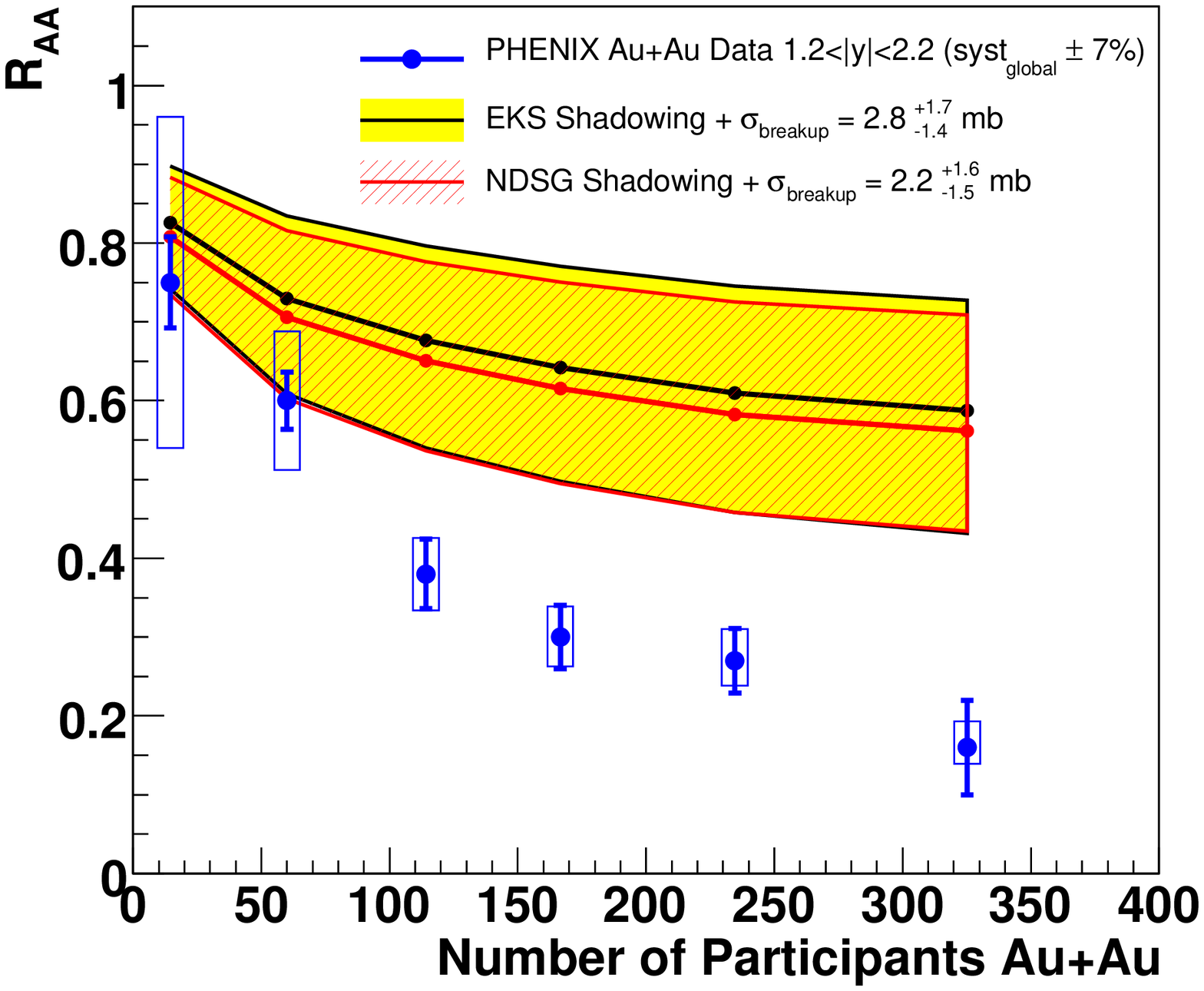}
  \vspace*{-1.2cm}
  \caption[]{Extrapolation of the simple CNM model shown in
  Figure \ref{rdau_eksmodel} to Au+Au collisions at forward rapidity. Results for
  both EKS\cite{EKS} and NDSG\cite{NDSG} shadowing are shown.}
  \label{figure_rauau_project_forw}
\end{minipage}
\vspace*{-0.8cm}
\end{figure}

In Figure \ref{rdau_eksmodel} an approximate constraint using a simple CNM model (with
shadowing and dissociation)\cite{vogt} is shown. This model can then be used to give an extrapolated
constraint for $Au+Au$ collisions, as shown in Figures \ref{figure_rauau_project_mid}
and \ref{figure_rauau_project_forw}. Clearly the $d+Au$ data from 2003 used to constrain
the CNM extrapolation here suffers from large uncertainties, and results in a large uncertainty
for $Au+Au$ collisions. For $Au+Au$ at mid-rapidity the CNM band is almost consistent with
the observed suppression - except for the most central collisions ($n_{part} \sim {340}$); while
at forward rapidity the suppression seen for $Au+Au$ is substantially stronger. The just
completed 2008 $d+Au$ run has approximately 30 times more $J/\psi$'s than before and, once
analyzed, will dramatically improve the knowledge of the CNM baseline in $A+A$ collisions,
and allow precision studies of the additional physics beyond CNM that comes from the hot-dense
matter created in heavy-ion collisions. The CNM constraint is expected to narrow by approximately
a factor of three with the new $d+Au$ data.

\section{How does the QGP affect Quarkonia}\label{QGP}  

\begin{figure}[htb]
\begin{minipage}[t]{0.50\linewidth}
  \includegraphics[width=\textwidth,clip=]{figures/raa_ratio_data.eps}
  \vspace*{-1.1cm}
  \caption[]{Nuclear modification factor for Au+Au collisions at mid
  rapidity (red circles), and at forward rapidity (blue squares) versus
  centrality (top panel). In
  the bottom panel the ratio of the foward over mid rapidity nuclear
  modification factors from the upper panel is shown.}
  \label{raa_ratio_data}
\end{minipage}
\hspace{0cm}
\begin{minipage}[t]{0.50\linewidth}
  \includegraphics[width=\textwidth,clip=]{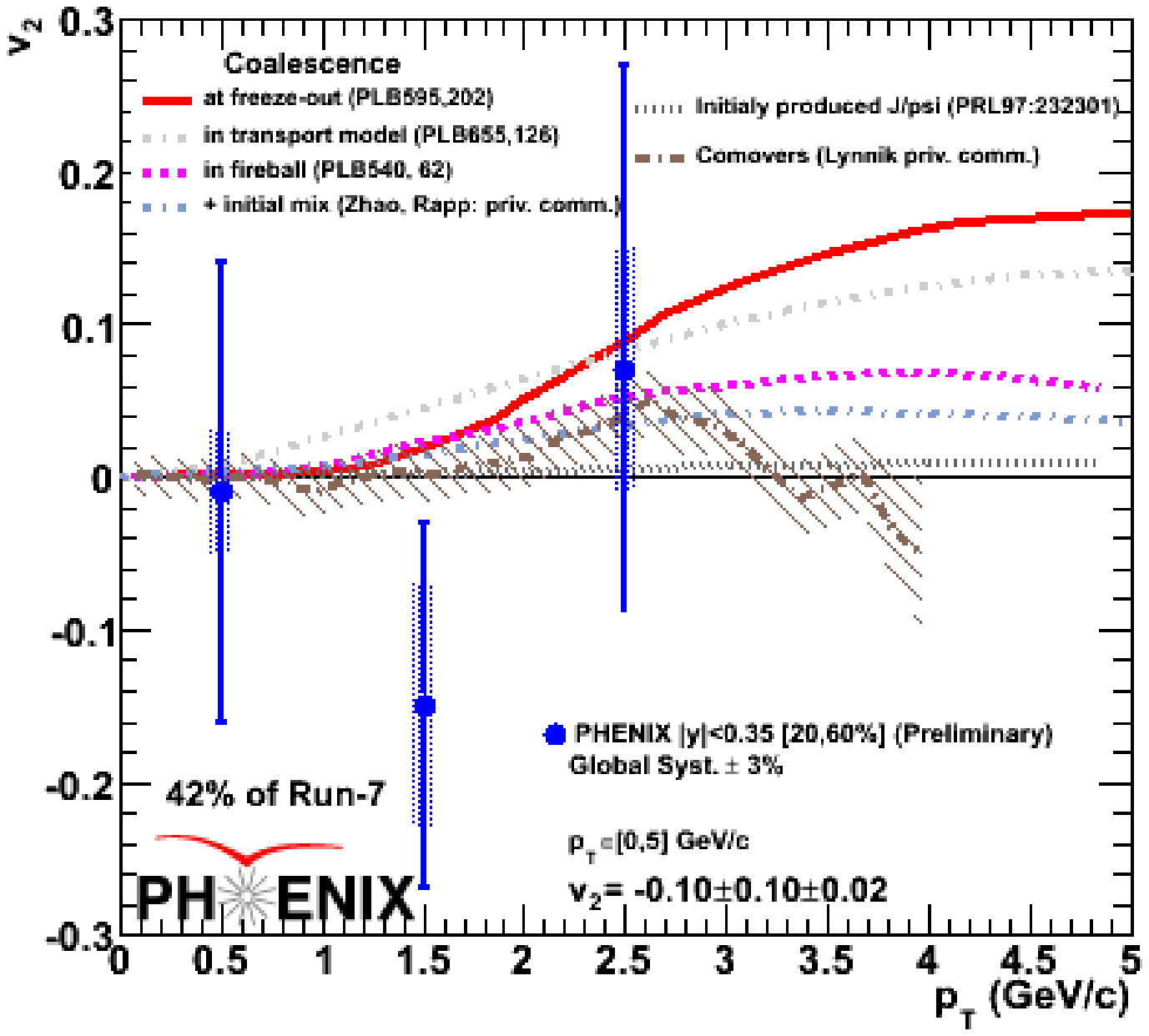}
  \vspace*{-1.2cm}
  \caption[]{Flow of $J/\psi$s at mid rapidity vs $p_T$
  (preliminary result from 42\% of the 2007 data), compared
  to several theoretical models.}
  \label{v2_pt_central_20-60_theories_prelim}
\end{minipage}
\vspace*{-0.8cm}
\end{figure}

Quarkonia are thought to be a definitive probe of the QGP through the
screening process in the deconfined colored medium\cite{satz}. Different quarkonia
states, because of their different binding energies, are expected to "melt" at
different temperatures of the medium. E.g. in some lattice calculations the $J/\psi$
would melt at $1.2 T_C$, but the $\Upsilon$ only at over $2 T_C$. Nuclear modification
factors observed by PHENIX in $Au+Au$ collisions are shown in Figure \ref{raa_ratio_data}.
The suppression at mid-rapidity is about the same as that observed for lower
energies at the SPS\cite{SPS}, despite the expectation that the hotter medium created at
RHIC would cause a larger suppression. The suppression at forward rapidity
is stronger than that at mid rapidity, and the ratio of the nuclear modification
factors, forward/mid, shown in the bottom panel of the figure, reaches an approximately
constant level of $0.6$ for $n_{part}>100$.

Several scenarios can be considered in trying to understand the observed trends:
1) CNM effects, as discussed above, should always be accounted for as a baseline.
2) Sequential screening\cite{screening} - where, as suggested by some lattice calculations,
only the $\psi^\prime$ and $\chi_C$ are screened and the $J/\psi$ itself is not - not at RHIC
or at SPS energies. Then the observed suppression beyond CNM comes only from loss
of the feeddown ($\sim{40\%}$) from the two higher mass quarkonia states.
3) Regeneration models\cite{regeneration},
where the large density of charm quarks created in the
collisions ($\sim 20$ in a central $Au+Au$ collision) can produce charmonia in the
latter stages of the expansion.

In the sequential screening picture, if the CNM suppression at mid rapidity
and the "melting" of the higher mass charmonia states was the same at RHIC and
at the SPS, this would provide a natural explanation for the nearly identical
suppression at RHIC and the SPS. It would also agree with some lattice
calculations that indicate no melting of the $J/\psi$ until over $2T_C$\cite{lattice_2tc}.
The stronger forward rapidity suppression seen at RHIC could then be explained by
gluon saturation that gives stronger forward suppression than that from standard
shadowing models.
For traditional shadowing models
the shadowing of the gluon from one nucleus is largely canceled by the anti-shadowing
from the gluon from the other nucleus - resulting in an approximately flat
rapidity dependence. For gluon saturation a "shadowing-like" effect is produced
for the gluon in the small-x region,
but no anti-shadowing for the other gluon, resulting in a stronger suppression at forward
rapidity. Since screening and gluon saturation might have different centrality dependences, it
is unclear whether they would balance to produce the approximately flat ratio
observed for $n_{part} > 100$ (Figure \ref{raa_ratio_data}).

An alternative is the regeneration picture, where the
dissocation by the QGP at mid and forward rapidity would be similar, but the
weaker suppression at mid rapidity would be due to regeneration effects
being stonger here where the charm density is largest. In this case it would
be an "accidental" compensation of screening and regeneration that leads to
the same mid-rapidity suppression at RHIC and the SPS.
At forward rapidity, where the charm density is smaller, the regeneration is
reduced and stronger screening results.
Again, whether the saturation in the forward to mid rapidity suppression could
be reproduced by these two compensating effects is unclear.

The regeneration mechanism depends on the square of the open-charm cross section,
so it is critical to resolve the present uncertainties there.\cite{open_charm}
Also, since charm has been shown to exhibit flow for moderate $p_T$ values,
one would expect $J/\psi$s that are produced by regeneration to inherit this
flow. A first measurement of the $J/\psi$ flow at mid rapidity is shown from part
of the 2007 $Au+Au$ data in Figure \ref{v2_pt_central_20-60_theories_prelim};
but is clearly quite challenging, and so far is consistent with zero flow.

\section{Summary and Future}\label{summary}  
The suppression of $J/\psi$ production in $Au+Au$ collisions at RHIC for mid rapidity is
similar to that at lower energies, while for foward rapidity the RHIC suppression is
stronger. Better cold nuclear matter constraints from the new $d+Au$ data are needed
to establish an accurate baseline and allow quantitative analysis of the QGP effects.
Two theoretical pictures,
1) sequential suppression with gluon saturation and
2) dissociation and regeneration, appear to offer explanations of the observed trends.
Higher luminosities and silicon vertex upgrades will enable much more quantitative
studies in the next few years. Over 100,000 $J/\psi$s and 600 $\Upsilon$s are expected
in a year with higher luminosities enabled by accelerator advances,
while new silicon vertex detectors will allow explicit indentification of
open-heavy and will improve both the background and mass resolution for the quankonia
states - especially important to separate the $\psi^\prime$ from the $J/\psi$ at forward
rapidity.

\vfill\eject

\begin{thebibliography}{9}  

\bibitem{e866_ups_pol} T. Chang {\it et al.}. (E866/NuSea), {\it Phys. Rev. Lett.}
  {\bf 91} (2003) 211801.

\bibitem{cdf_polarization} T. Affolder et al. (CDF), {\it Phys. Rev. Lett.}  {\bf 85} (2000) 2886.

\bibitem{lansberg} H. Haberzettl and J.P. Lansberg, {\it Phys. Rev. Lett.}  {\bf 100} (2008) 032006.

\bibitem{jpsi_run5pp} A. Adare {\it et al.}, (PHENIX), {\it Phys. Rev. Lett.}
  {\bf 98} (2007) 232002.

\bibitem{moreno} Y. Morino (PHENIX), this proceedings.

\bibitem{ppg078} A. Adare {\it et al.}, (PHENIX), {\it Phys. Rev.} {\bf C77} (2008) 024912.

\bibitem{EKS} K.J. Eskola, V.J. Kolhinen, and R. Vogt, {\it Nucl. Phys.} {\bf A696} (2001) 729.

\bibitem{NDSG} D. deFlorian and R. Sassot, {\it Phys. Rev.} {\bf D69} (2004) 074028.

\bibitem{CGC} L. McLerran and R. Venugopalan, {\it Phys. Rev.} {\bf D49} 2233 (1994);
3352 (1994).

\bibitem{vogt} R. Vogt, {\it Phys. Rev.} {\bf C77} (2005) 054902.

\bibitem{satz} T. Matsui and H. Satz, {\it Phys. Lett.} {\bf B178} (1986) 416.

\bibitem{SPS} M.C. Abreu {\it et al.}. (NA50) {\it Phys. Lett.} {\bf B477} (2000) 28; Phys. Lett. B521 (2001)
195.

\bibitem{screening} F. Karsch, D. Kharzeev, H. Satz, {\it Phys. Lett.} {\bf B637} (2006) 75;
hep-ph/0512239.

\bibitem{regeneration} L. Grandchamp, R. Rapp, G.E. Brown, {\it Phys. Rev. Lett.} {\bf 92} (2004) 212301;
R.L. Thews, {\it Eur. Phys. J} {\bf C43} (2005) 97.

\bibitem{lattice_2tc} F. Datta {\it et al.}, hep-lat/0409147.

\bibitem{open_charm} A. Knospe (STAR), this proceedings.
  
\end{thebibliography}
\end{document}